# Reactive sputtering of SnS thin films using sulfur plasma and a metallic tin target: achieving stoichiometry and large grains

Daiki Motai[1], Issei Suzuki[1,*], Taichi Nogami[1], Takahisa Omata[1]

1. Institute of Multidisciplinary Research for Advanced Materials, Tohoku University, Sendai 980-8577, Japan

*Corresponding author: issei.suzuki@tohoku.ac.jp

**Abstract**

This study presents a novel method for fabricating stoichiometric SnS thin films with large grain sizes via reactive sputtering using a metallic Sn target and sulfur plasma (S-plasma). Unlike conventional approaches that rely on toxic $H_2S$ gas, this method employs a S-plasma to enhance sulfur reactivity and mitigate sulfur deficiencies during film deposition. By optimizing the balance between the sputtering conditions of the Sn target and the supply conditions of the S-plasma, dense single-phase SnS thin films with micron-scale grain sizes were achieved at a substrate temperature of 300 °C, achieving an in-plane Hall mobility of 13 $cm^2$ $V^{-1}$ $s^{-1}$. Furthermore, crystalline SnS thin films were fabricated even on a room-temperature substrate, enabling potential applications in flexible devices with heat-sensitive substrates. These findings demonstrate the effectiveness of S-plasma in advancing SnS thin film fabrication, providing a safer and more efficient route to high-performance photovoltaic materials.

## Introduction

Orthorhombic α-SnS (SnS) is a compound semiconductor composed of abundant and non-toxic elements. SnS exhibits an indirect band gap of 1.1 eV and a direct band gap of 1.3 eV.[1] Due to the small energy difference between its direct and indirect band gaps, SnS exhibits a large light absorption coefficient exceeding $1 \times 10^4$ cm$^{-1}$ above its band gap energy.[2] Because acceptor-type intrinsic defects (e.g., Sn vacancies ($V_{Sn}$) and Sn anti-sites ($Sn_S$)) are easily generated in SnS, it generally exhibits p-type electrical conduction. Since the pioneering work by Noguchi et al. in 1994,[3] SnS has been actively studied as a promising material for the light-absorbing p-type layer of thin-film solar cells.[3–5] However, the power conversion efficiency (PCE) of SnS thin-film solar cells has stagnated at a maximum of 4.8%.[3,6] One of the factors limiting the PCE of SnS solar cells is the formation of defects caused by sulfur deficiencies.[7] Sulfur deficiencies lead to the formation of donor-type sulfur vacancies ($V_S$) and acceptor-type defects, such as $Sn_S$ and $V_{Sn}$, due to the migration of Sn into the $V_S$. These defects induce Fermi level pinning, preventing the splitting of quasi-Fermi levels, which is essential for the origin of open-circuit voltage ($V_{OC}$), thereby reducing the $V_{OC}$. Therefore, suppressing sulfur deficiencies is crucial for improving the PCE of SnS thin-film solar cells.

Conventional measures to suppress sulfur deficiencies in SnS thin films follow two main approaches. The first approach involves the depositions of thin films in an environment with a high sulfur partial pressure using sputtering or pulsed laser deposition (PLD), while the second approach compensates for sulfur deficiencies through post-annealing after the fabrication of sulfur-deficient thin films. Sulfur vapor primarily consists of stable, crown-shaped $S_8$ molecules, which have inherently low reactivity.[8] Consequently, $H_2S$ has been conventionally used as a sulfur source for these approaches.[9,10] However, the use of $H_2S$ leads to the incorporation of hydrogen impurities into SnS thin films, inevitably forming acceptor-type defects ($H_{Sn}$)[11] (in the Kröger–Vink notation, $V_S^{\cdot\cdot} + 2e' + H_2S \rightarrow S_S^{\times} + 2H_{Sn}' + 2h^{\cdot}$). Furthermore, the use of toxic $H_2S$ is undesirable due to concerns regarding process safety and management costs. Suzuki et al.[12] and Nogami et al.[13] have proposed a process for suppressing sulfur deficiencies by supplying sulfur plasma (S-plasma) during sputtering with a SnS sintered target. In this process, it is understood that the application of a radio frequency (RF) to sulfur vapor leads to the cleavage of S–S bonds (i.e., the cracking of $S_8$ molecules), thereby generating highly reactive sulfur species (atoms, molecules, and ions). As mentioned earlier, sulfur vapor ($S_8$) has low reactivity, and to date, there have been no reports on reactive sputtering using a metallic target and sulfur vapor. By utilizing highly reactive S-plasma, sulfides are expected to form through reactive sputtering with metallic targets.

In this study, SnS thin films were fabricated by reactive sputtering using a metallic Sn target and S-plasma. In the conventional sputtering process using a compound sintered target, it is known that the target composition changes over time during sputtering due to differences in the sputtering rates and vapor pressures of its constituent elements.[14,15] In contrast, a metallic target has the advantage of being free from compositional changes over time during sputtering. This study demonstrates that stoichiometric, single-phase SnS thin films can be fabricated by optimizing the balance between Sn and S supply. In addition, this method enables the fabrication of dense SnS thin films composed of micron-scale grains, which could not be achieved by conventional sputtering using a SnS target. These films exhibit high in-plane hole mobility (13 cm$^2$ V$^{-1}$ S$^{-1}$). Furthermore, this method demonstrated that crystalline SnS thin films can be fabricated without substrate heating due to the high reactivity of S-plasma. Because crystalline SnS thin films can also be deposited on heat-sensitive polymer substrates, this approach is expected to be applicable to flexible devices.

## Experimental

SnS thin films were fabricated on 40 × 40 mm$^2$ SiO$_2$ glass substrates using deposition system consisting of a sputtering cathode with a metallic tin target and a S-plasma source (Figure 1(a)). The substrates were either not intentionally heated ($T_{Sub}$ = room temperature (RT)) or heated to 300 ºC ($T_{Sub}$ = 300 °C) using a substrate heater. The base pressure in the deposition chamber was on the order of 10$^{-6}$ Pa. A 1-inch metallic Sn target (99.99%) was sputtered using a pressure-gradient sputtering cathode (Kenix, Co., Ltd., Japan), while S-plasma was introduced to the thin-film deposition area using a S-plasma source (RF plasma cracking cell; Kenix Co., Ltd.). As shown in Figure1(b), sulfur powder is heated to 100 °C at the bottom to generate sulfur vapor, which is then converted into S-plasma by applying RF power (50 W) with inductively coupled plasma (ICP) coil and introduced into the deposition chamber along with Ar carrier gas (5 sccm).[13] The distance from the top of S-plasma source to the substrate was fixed at 65 mm, while the distance from the Sn target to the substrate surface (*T-S*) was set to either 65 or 135 mm. The RF power applied to the Sn target ($RF_{Sn}$) was set to 5 or 10 W, and the deposition time was 5 h. The phase of fabricated thin films was identified by X-ray diffraction (XRD; CuKα line, SmartLab, Rigaku Corp., Japan). The composition of the thin films was determined using a field-emission electron probe microanalyzer (FE-EPMA, JXA-8530, JEOL Ltd., Japan). The structure of the thin films were investigated using Raman spectroscopy (inVia, Renishaw plc., UK). The surface and cross-sectional morphology and film thickness were observed using a Schottky field-emission scanning electron microscope (FE-SEM; JSM-7800F, JEOL Ltd., Japan). Electrical properties were evaluated at RT by measuring the electrical conductivity via the van der Pauw method and Hall effect measurements (ResiTest 8300, TOYO Corp., Japan).

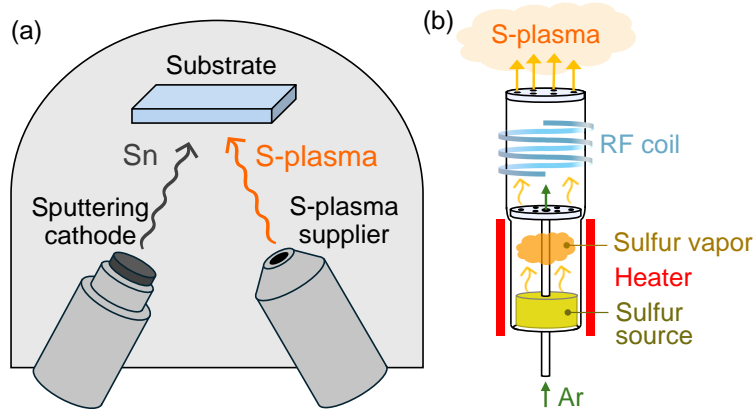

Figure 1. (a) Schematic of the reactive sputtering system composed of a sputtering cathode and a S-plasma supplier. (b) Detailed structure of the S-plasma supplier. Note that the scale does not necessarily reflect the actual objects accurately.

## Results

Figure 2(a) shows the XRD profiles of three thin films fabricated under different Sn sputtering conditions at $T_{Sub}$ = RT with S-plasma supply. A lower $RF_{Sn}$ and longer $T$-$S$ corresponds to a lower supply of Sn to the deposition area. At $T$-$S$ = 65 mm and $RF_{Sn}$ = 10 or 5 W, the fabricated thin films were a mixture of β-Sn and SnS; this result is consistent with the S/(S+Sn) of the thin films, which are 0.22 and 0.29, respectively, both lower than stoichiometric composition. With $RF_{Sn}$ = 5 W and increasing $T$-$S$ to 135 mm resulted in the deposition of single-phase SnS. The composition of this thin film was determined to be S/(Sn+S) = 0.5, which corresponds to the stoichiometric composition. A small amount of impurity phases, such as $SnS_2$ or $Sn_2S_3$, are often detected by Raman spectroscopy in SnS thin films. However, in the SnS thin films fabricated in this study, the peak at ~310 cm$^{-1}$, which is attributed to the $Sn^{4+}$–S bond, was not observed, indicating that the SnS thin films were free of different phases (Figure 3(a)). The supply of Sn and S species to the deposition area, as well as the amount actually deposited as a thin film (i.e., material balance), were estimated based on the deposition rate of metallic Sn and the weight loss of sulfur source from the S-plasma supplier. The majority (~70%) of Sn atoms supplied from the sputtering cathode were deposited as SnS thin films, while only 1–2% of the sulfur supplied from the S-plasma source was incorporated into the thin films (for a detailed analysis, see Section S1 in the Supplementary Material). This indicates that the environment near the substrate surface during deposition is highly rich in reactive sulfur species. In addition, when the same amount of sulfur vapor ($S_8$) was supplied instead of S-plasma (i.e., sulfur powder was heated to evaporate but without the application of RF power), the resulting phase was almost entirely metallic β-Sn (Figure 2(b)). This indicates that in the reactive sputtering of Sn and S,

SnS thin film forms not only due to the excessive supply of sulfur but also because sulfur is supplied in a highly reactive plasma state.

The XRD profile of the fabricated SnS thin films (bottom of Figure 2(a)) exhibited strong peaks that may be attributed to the 200, 400, and 800 diffraction peaks. Because the intensities of 400 and 111 diffraction peaks of the powder pattern are both strong, and their diffraction angles are close, careful attention is required when determining the presence of the 111 diffractions.[13] In this study, as the three main peaks fit the 200, 400, and 800 diffraction peaks from the same lattice parameter, it is clear that the XRD spectrum does not include the 111 diffraction peak (for a detailed analysis of the XRD profiles, see Figure S2 in the Supplementary Materials). However, in the in-plane XRD profile, diffraction attributed to the planes not perpendicular to {$h$00} plane were also observed (Figure S3 in the Supplementary Material), indicating that the thin films also contain crystallites with partially tilted orientations. SnS has a layered structure with the {$h$00} plane which acts as the Van der Waals plane, and which has the lowest surface energy.[16,17] This potentially rationalizes why the SnS thin films exhibit self-oriented growth even when deposited on amorphous $SiO_2$ glass substrates, aligning with other studies.[18]

Figure 2(b) shows the XRD profile and composition of the SnS thin film deposited at $T_{Sub}$ = 300 °C. Similar to the thin film fabricated at $T_{Sub}$ = RT, the 200, 400, and 800 diffraction peaks were observed, and the film composition was stoichiometric. The observed peaks are clearly sharper than those of the thin film fabricated at $T_{Sub}$ = RT. As shown in Figure 4, the full widths at half maximum (FWHMs) of the 400 diffractions of the obtained SnS thin films decreased steeply with increasing $T_{Sub}$, and became significantly narrower at $T_{Sub}$ = 300 °C than those obtained through deposition or annealing at above 300 °C in other techniques. Although the FWHM of diffraction peaks is influenced by multiple factors—including crystallite size, strain, and preferential orientation—and serves as an indicator of what is often loosely referred to as "crystallinity", making quantitative evaluation challenging, these results suggests that the thin films obtained in this study are generally superior in these aspects. In addition, in the Raman spectrum (Figure 3), the peaks of the thin film fabricated at $T_{Sub}$ = 300 °C are significantly sharper than those of the thin film fabricated at $T_{Sub}$ = RT , which is consistent with the peak narrowing observed in the XRD profiles.

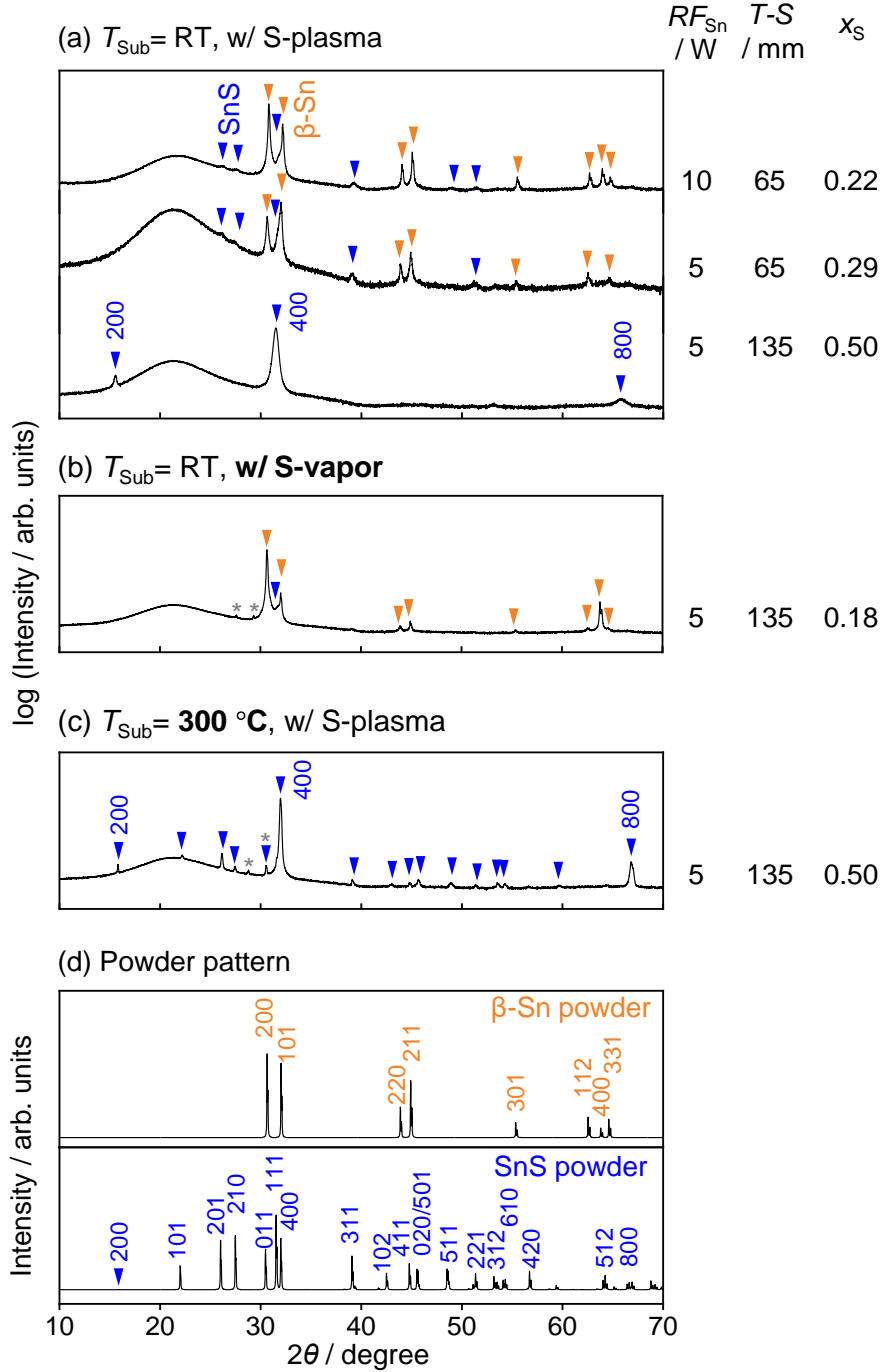

Figure 2. XRD profiles of the thin films deposited at (a) $T_{sub}$ = RT and (b) 300 °C with S-plasma supply, and (c) at $T_{sub}$ = RT with S-vapor. (d) Powder patterns of β-Sn [19] and α-SnS [20] provided for comparison. The sputtering condition of $RF_{Sn}$, $T$-$S$, and the resulting composition, $x_S$ = S/(Sn+S), are denoted on the right-hand side. Note that the broad peak around 21° is a halo pattern due to the $SiO_2$ substrate. The peaks marked with an asterisk are produced by Cu Kβ and W Lα radiations.

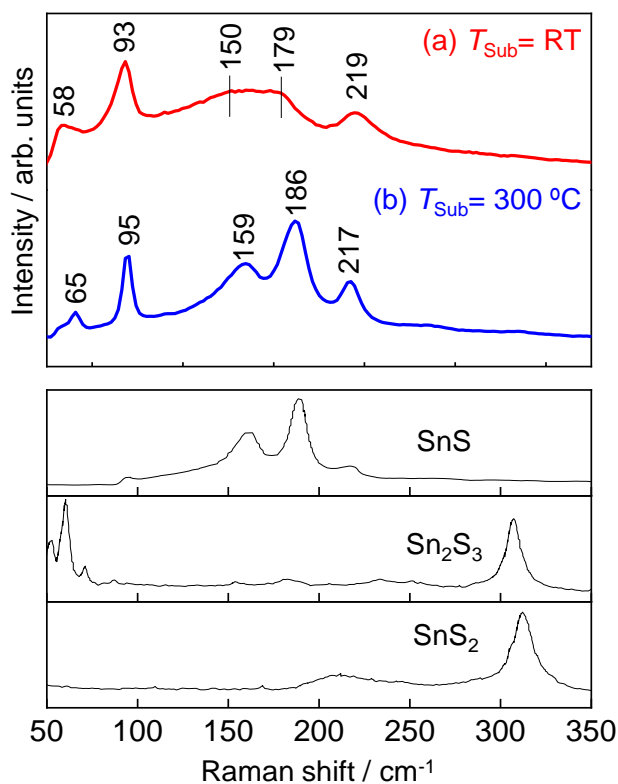

Figure 3. Raman spectra of SnS thin films deposited at (a) $T_{Sub}$ = RT and (b) 300 °C with $RF_{Sn}$ = 5 W and $T$-$S$ = 10 mm. The spectra of SnS, $Sn_2S_3$, and $SnS_2$ [21,22] are shown for comparison.

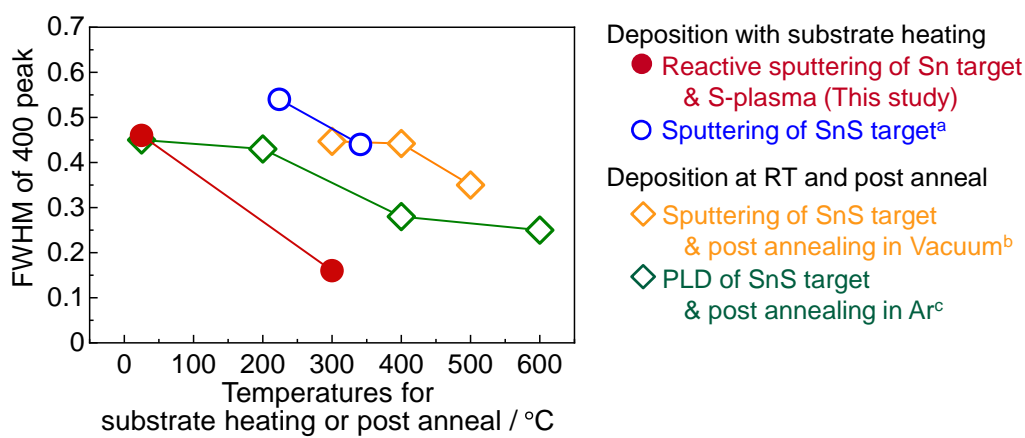

Figure 4. FWHM of 400-diffraction peaks of SnS thin films in XRD profiles fabricated in this study, compared with those previously reported in the literature, which also exhibited self-preferential $h$00 orientation, as a function of substrate heating or post-annealing temperature. Literature values are cited from [a]Ref.[12], [a]Ref.[23], [c]Ref.[18].

Figure 5 shows the SEM images of the surface and cross-section of the thin films. Grain boundaries were not clearly observed for the thin films fabricated at $T_{Sub}$ = RT, indicating that the thin film was composed of significantly small grains. In contrast, thin films fabricated at $T_{Sub}$ = 300 °C consist of large grains of over 1 μm and relatively small grains of several hundred nm. The surface of the large grains clearly exhibited facets, indicating that these grains are single crystals. From the cross-sectional SEM image, the thin films exhibited a dense morphology with no grain boundaries in the thickness direction within the large grains. Such a morphology of thin films is preferred for photovoltaic applications, because carriers transport in the thickness direction in the devices, and fewer grain boundaries suppress carrier recombination. Therefore, this SnS thin film is highly suitable for application in SnS thin-film solar cells.

The electrical properties SnS thin films obtained in this study are summarized at Table 1, and Figure 6 shows a comparison of their Hall mobilities and carrier densities with those previously reported for SnS thin films fabricated by sputtering. The SnS thin films fabricated in this study exhibit a carrier density on the order of $10^{16}$ cm$^{-3}$. Given that undoped SnS thin films fabricated by various methods, including sputtering, PLD, and evaporation, typically show carrier densities in the range from $10^{15}$ to $10^{16}$ cm$^{-3}$,[24,25] the carrier density of the thin films obtained in this study is within the typical range for SnS thin films. The hole mobility of SnS fabricated at $T_{Sub}$ = RT was 2.3 cm$^2$ V$^{-1}$ s$^{-1}$, which improved to 13 cm$^2$ V$^{-1}$ s$^{-1}$ at $T_{Sub}$ = 300 °C. Although not as high as the mobility of undoped-SnS single crystals (~90 cm$^2$ V$^{-1}$ s$^{-1}$),[26] this value is significantly high for SnS thin films fabricated by sputtering. It has been previously reported that the carrier transport in SnS thin films is dominated by grain boundary scattering.[9,12,27] The high Hall mobility obtained in this study is therefore attributed to the reduced grain boundary density due to the large grain size.

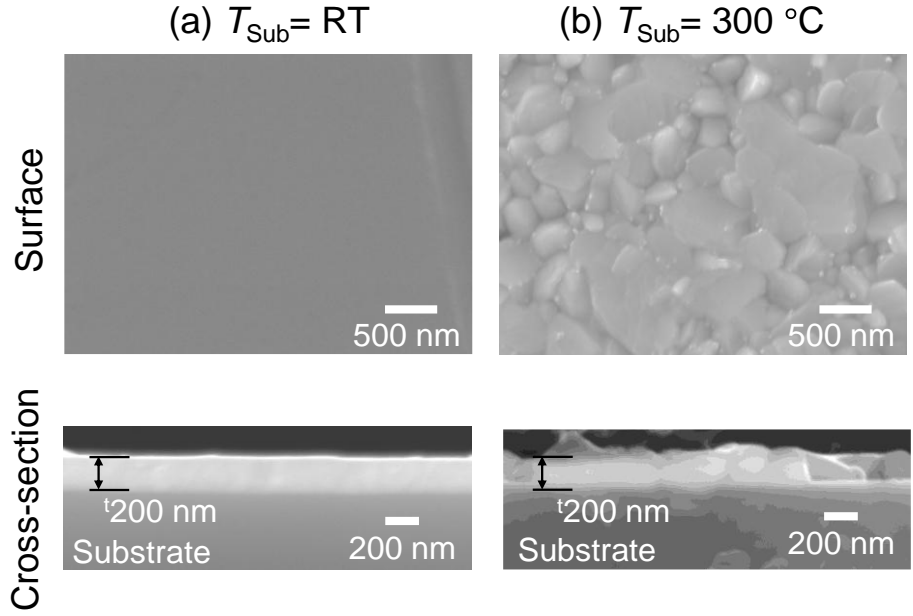

Figure 5. SEM images of the surface and cross-section of SnS thin films deposited at (a) $T_{Sub}$ = RT and (b) 300 °C Note that with $RF_{Sn}$ = 5 W and $T$-$S$ = 135 mm.

Table 1. Electrical properties of the SnS thin films deposited at $T_{sub}$ = RT and 300 °C determined by Hall measurement, along with those of undoped SnS single crystal. Note that $RF_{Sn}$ and $T$-$S$ were 5 W and 135 mm, respectively.

|  | $T_{Sub}$ = RT | $T_{Sub}$ = 300 °C | SnS single crystal[a] |
|---|---|---|---|
| Carrier type | p-type | p-type | p-type |
| Electrical conductivity, σ / S·cm$^{-1}$ | $4.2 \times 10^{-3}$ | $6.9 \times 10^{-2}$ | $1.8 \times 10^{-1}$ |
| Carrier density, $n$ / cm$^{-3}$ | $1.2 \times 10^{16}$ | $3.4 \times 10^{16}$ | $1.3 \times 10^{16}$ |
| Hole mobility, $\mu$ / cm$^2 \cdot$V$^{-1} \cdot$s$^{-1}$ | 2.3 | 13 | 91 |

[a]Ref. [26]

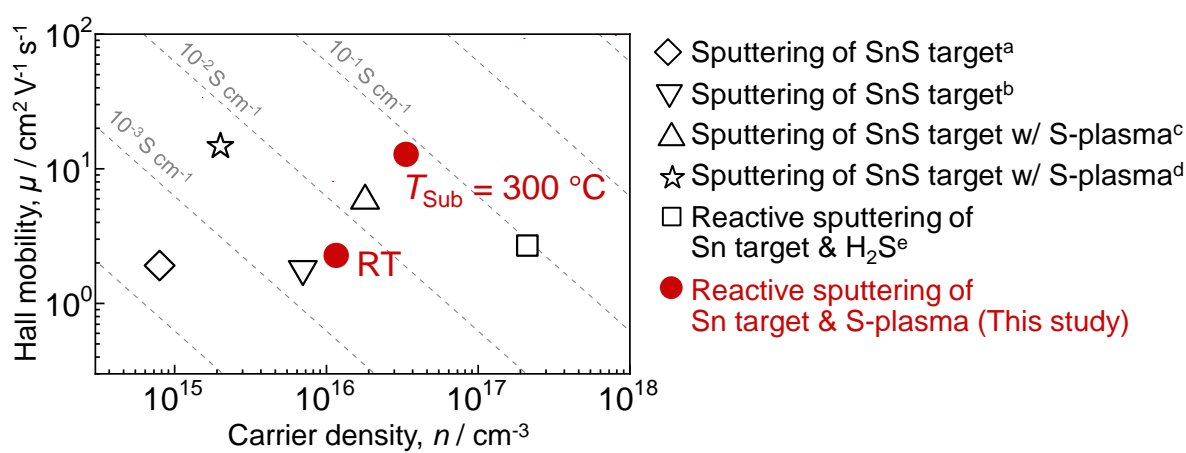

Figure 6. Comparison of the Hall mobility and carrier density of SnS thin films fabricated by the sputtering technique in this study and those of previously reported studies: [a]Ref. [12], [b]Ref. [28], [c]Ref. [29], [d]Ref. [13], and [e]Ref. [30].

## Discussions

It is generally known that it is significantly difficult to fabricate SnS thin films with a morphology consisting of large grains. For example, in reactive sputtering using a Sn target and $H_2S$ gas, the resulting SnS thin film consists of plate-like grains, with diameters of several tens of nanometers at $T_{Sub}$ = 300 °C and 100–500 nm at $T_{Sub}$ = 400 °C.[30] Furthermore, sputtering with a SnS target results in the formation of SnS grains with diameters ranging from 100 to 200 nm at $T_{Sub}$ between RT and 300 °C.[12,31] In general, increasing $T_{Sub}$ is an effective approach for increasing grain size. For example, for the deposition of oxide semiconductors, a high $T_{Sub}$ of up to 1000 °C is often employed to increase grain size. However, this is not the case for SnS, because when SnS is deposited by PVD, re-evaporation of SnS occurs at approximately 400 °C,[32] making it impossible to use an higher $T_{Sub}$ to induce grain growth. Post-annealing in a $H_2S$ atmosphere or vacuum has been employed as a means of increasing the grain size of SnS.[33,34] Nevertheless, significant crystal grain growth has not been achieved. To the best of our knowledge, a sputtering method for achieving large grains on the micron scale has not yet been reported.

Here, we discuss why the reactive sputtering using S-plasma in this study promoted grain growth up to the micron scale. According to the extended structure zone diagram for sputtering by Anders, a high sum of the thin-film homologous temperature $T_h$ and the potential temperature $T_p$ (which is determined by the potential energy of particles reaching the substrate) enhances atomic rearrangement and promotes grain growth.[35] In this study, significant grain growth was observed even at low $T_{Sub}$, indicating that $T_p$ was increased by supplying S-plasma. It has been reported that when RF is applied to sulfur vapor with ICP coil, $S_8$ molecule dissociates into atomic species and further ionized into S⁺ and S⁻.[36,37] As shown in Figure 7, S and S⁺ possess extremely high chemical potential. Therefore, when these particles reach the substrate and contribute to SnS formation, it should increase $T_p$ leading to grain growth. Additionally, as described at the beginning of the Results section, only 1–2% of the sulfur supplied was incorporated into the SnS thin film, indicating that the majority of sulfur species reaching the substrate does not remain. It is well known that ions not incorporated into the thin film neutralize on the film surface while supplying electronic energy and then leave.[35] In this study, if S⁻ and S⁺ are neutralized on the film surface and leave as neutral S, the energy supplied to the film corresponds to approximately 7 eV·particle⁻¹ (sum of the electron affinity of S, ~2 eV, and the work function of SnS, ~5 eV [26]) and approximately 5 eV·particle⁻¹ (difference between the ionization energy of S, ~10 eV, and the work function of SnS), respectively, which also contributes to the increase in $T_p$. Based on these considerations, it is supposed that the high chemical potential of S-plasma and the supply of electronic energy by S ions not incorporated into the film are the origin of the significant promotion of the grain growth. In contrast, it has also been reported that when Ar plasma bombards sulfide targets such as ZnS, Cu₂S, and FeS, the sulfur component is sputtered as neutral S₂ molecules, irrespective of the crystal structure of the target.[38] Because S₂ molecules have low chemical potential (see Figure 7) and do not supply electronic energy to the substrate surface, $T_p$ remains low, which likely prevents grain growth during conventional sputtering using SnS targets.

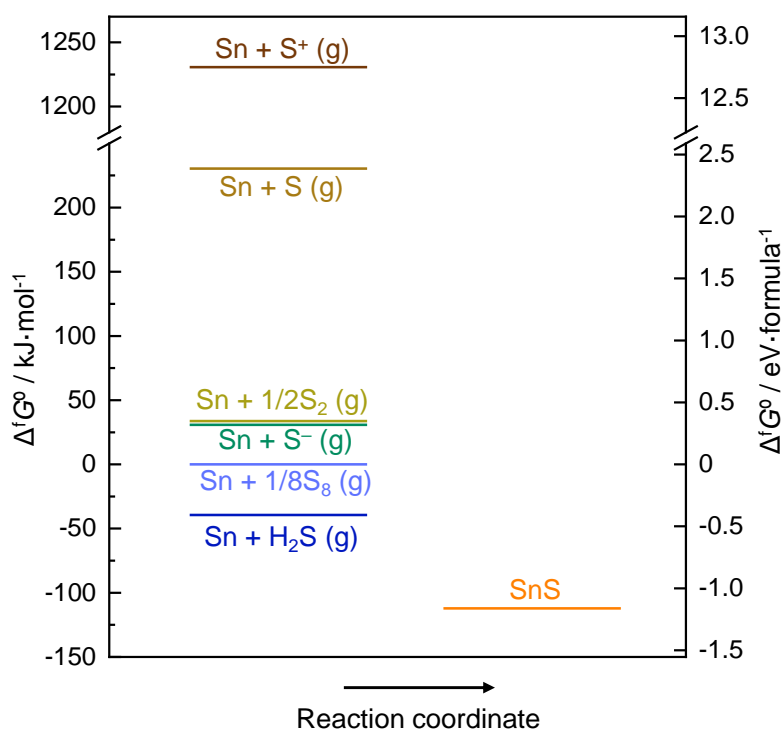

Figure 7. Thermodynamics of SnS formation at 298.25 K from various reactants.[39,40]

## Summary

This work demonstrates the effectiveness of reactive sputtering with S-plasma supply in achieving stoichiometric SnS thin films. Crystalline SnS films were obtained even at $T_{Sub}$ = RT, and micron-sized grains and high mobility were achieved at $T_{Sub}$ = 300 °C, which is significantly suitable for photovoltaic applications. The observed significant grain growth should be attributable to the high chemical potential of S-plasma and the electronic energy supplied to the growing thin-film surface by the excessive S ions not incorporated into the films. These results demonstrate that the fabricated SnS thin films possess desirable material properties for energy applications, particularly in photovoltaics, through improved crystallinity and carrier mobility.

**Acknowledgement**

The authors would like to thank Prof. Takeo Nakano at Seikei University for his invaluable advice on the hypothesis regarding the influence of S-plasma on the thin-film growth process. The authors appreciate the kind support of Dr. Takeo Ohsawa and Dr. Naoki Ohashi at the National Institute for Materials Science (NIMS) in the electrical property analysis using Hall measurements.

**Funding sources**

This work was partly supported by a Grant-in-Aid for Scientific Research (B) (Grant No. 21H01613), the Research Program of "Five-Star Alliance" in "NJRC Mater. & Dev.", and the Tsukuba Innovative Arena (TIA) "Kakehashi" collaborative research program. The authors thank Mr. Kei Toyooka at Tohoku University for his help with the cross-sectional SEM observations. T. N. was financially supported by JST SPRING, Grant Number JPMJSP2114.

**CRediT authorship contribution statement**

**Daiki Motai**: Investigation (lead), Visualization (lead), Writing – original draft (equal). **Issei Suzuki:** Conceptualization (lead), Investigation (supporting), Funding acquisition (lead), Writing – original draft (equal), Supervision (equal). **Taichi Nogami:** Investigation (supporting). **Takahisa Omata:** Writing – review & editing (equal), Supervision (equal).

**Supplementary Material**

See the Supplementary Material for details regarding the material balance of the reactive sputtering, enlarged XRD profiles, and in-plane XRD profiles.

**Declaration of competing interest**

The authors declare that they have no known competing financial interests or personal relationships that could have appeared to influence the work reported in this paper.

# Supplementary Material for

# Reactive sputtering of SnS thin films using sulfur plasma and a metallic tin target: achieving stoichiometry and large grains

Daiki Motai[1], Issei Suzuki[1,*], Taichi Nogami[1], Takahisa Omata[1]

1. Institute of Multidisciplinary Research for Advanced Materials, Tohoku University, Sendai 980-8577, Japan

*Corresponding author: issei.suzuki@tohoku.ac.jp

## Section S1. Material balance and incorporation efficiency in reactive sputtering

When metallic Sn thin films were deposited by sputtering using a Sn target ($RF_{Sn}$ = 5 W) without S-plasma or sulfur vapor, the deposition rate was 33 nm h$^{-1}$ (Figure S1). In this case, assuming a relative density of 100% for metallic Sn thin films and sticking probability of 100% for atoms reaching the substrate surface, the flux of Sn atoms reaching the substrate surface was estimated to be 1.2 × 10$^{17}$ atoms cm$^{-2}$ h$^{-1}$.

The thickness of the SnS thin films fabricated in this study was 200 nm regardless of $T_{Sub}$, corresponding to a deposition rate 40 nm h$^{-1}$. Based on the assumption that the relative density of the SnS thin film is 100%, the incorporated Sn atoms as SnS thin film was 8.4 × 10$^{16}$ atoms cm$^{-2}$ h$^{-1}$. Thus, the fraction of Sn atoms reaching the substrate surface and being incorporated into the thin film was estimated to be approximately 70%.

When sulfur powder in the sulfur plasma source was heated at 100 °C, the weight loss of sulfur powder (supplied to the chamber) was 3–7 mg h$^{-1}$.[S1] Based on the actual deposition rate of SnS (40 nm h$^{-1}$, as described above), the amount of sulfur incorporated into the SnS thin films on the 40 × 40 mm$^2$ substrate was 0.07 mg h$^{-1}$. Therefore, 1–2% of the S-plasma supplied was incorporated into the SnS thin films. It is assumed that most of the remaining S-plasma either deposited outside the substrate or was exhausted from the chamber.

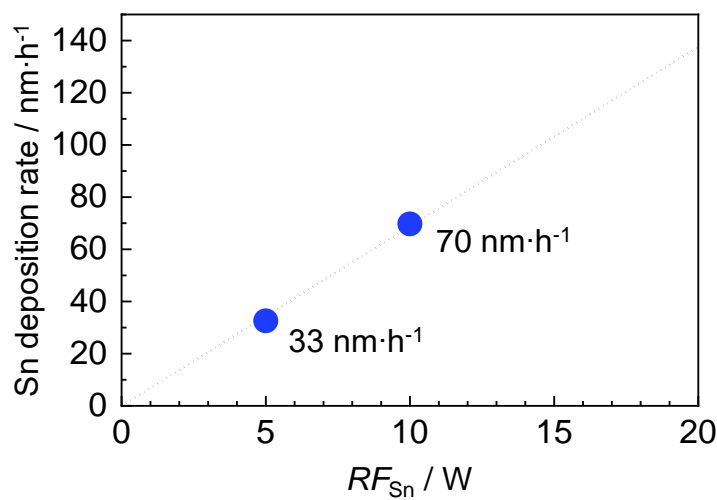

Figure S1. Deposition rate of metallic Sn as a function of RF power for a metallic Sn target deposited with *T-S* = 135 mm without S-plasma or sulfur vapor supply.

**Section S2. Enlarged XRD profiles of the SnS thin films**

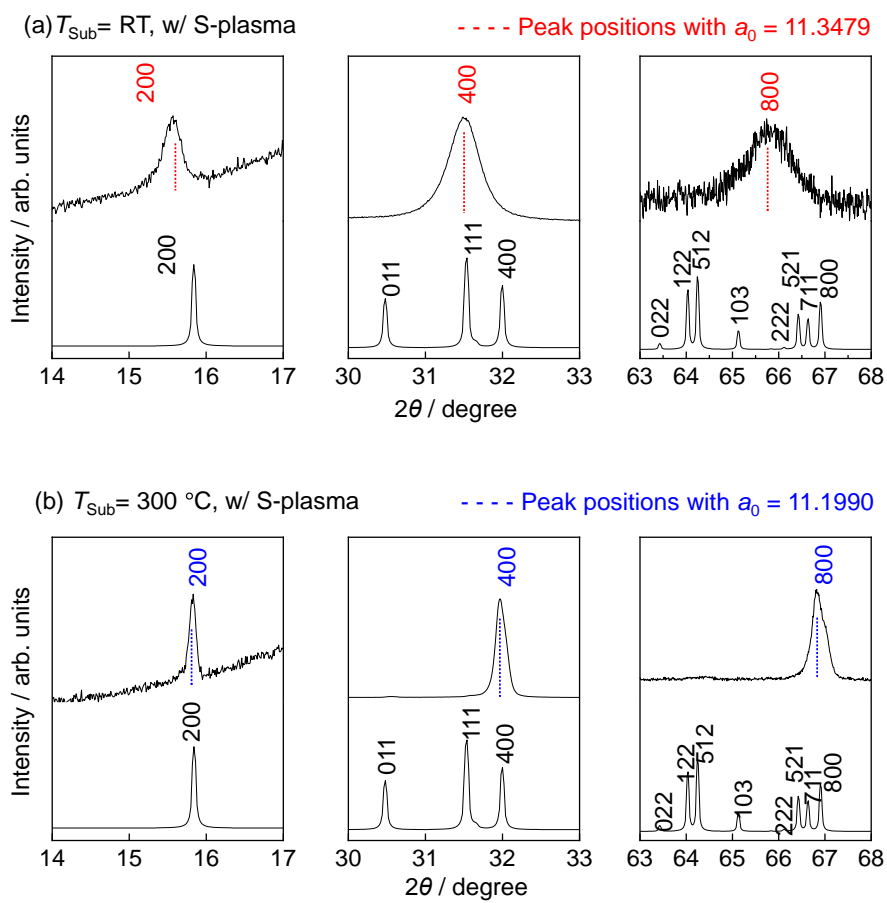

Figure S2. Enlarged XRD profiles of SnS thin films fabricated at (a) $T_{Sub}$ = RT and (b) 300 °C. The vertical dotted lines in blue or red indicate the corresponding peak positions for the assumed lattice parameter or lattice plane spacing, which are shown in the upper right corner of each figure.

## Section S3. In-plane XRD analysis

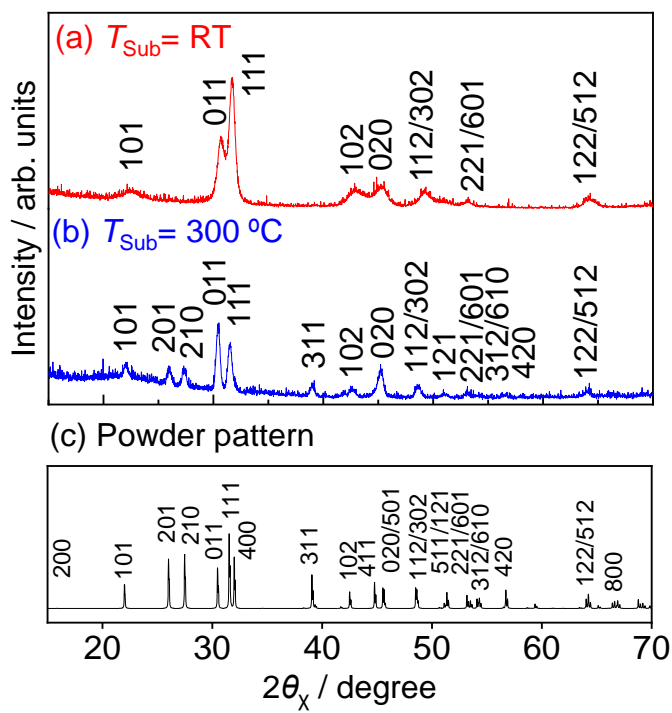

Figure S3. In-plane XRD profiles of thin films fabricated at (a) $T_{Sub}$ = RT and (b) 300 °C with $RF_{Sn}$ = 5 W and $T$-$S$ = 10 mm, along with the powder pattern of SnS (ICSD#24376).[S2]

## References for supplementary material